\documentstyle[12pt,twoside,fleqn,espcrc1,epsfig]{article}

\def\pl#1#2#3   {{\em Phys. Lett.} {\bf#1} (#2) #3}
\def\np#1#2#3   {{\em Nucl. Phys.} {\bf#1} (#2) #3}
\def\epj#1#2#3  {{\em Eur. Phys. J.} {\bf#1} (#2) #3}
\def\prl#1#2#3  {{\em Phys. Rev. Lett.} {\bf#1} (#2) #3}
\def\zp#1#2#3   {{\em Zeit. Phys.} {\bf#1} (#2) #3}

\newcommand{\Pom}{{I\!\!P}}
\newcommand{\Reg}{{I\!\!R}}

\newcommand{\qsq} {$Q^2$}

\newcommand{\xpom} {$x_\Pom$}
\newcommand{\ttt} {$t$}

\newcommand{\fdthree} {$F_2^{D(3)}$}
\newcommand{\fdthreef} {$F_2^{D(3)}  (Q^2, x_{I\!\!P} , \beta )$}
\newcommand{\fdtwo} {$F_2^{D(2)} (Q^2,\beta)$}

\title{Diffractive Physics at HERA}

\author{L. Favart\address{Charg\'e de recherche FNRS et 
        Universit\'e Libre de Bruxelles, 
        1050 Brussels, Belgium \\
        on behalf of the H1 and ZEUS Collaborations}}

\begin{document}
\maketitle

\section{Introduction}

Presently, one of the most important tasks in particle physics is the 
understanding of the strong force.
For this purpose, the Quantum Chromodynamics theory (QCD), 
part of the Standard Model, 
seems to be the best candidate.
An important characteristic of this theory is that the 
coupling constant $\alpha_S$ tends to zero when the transverse distance 
between matter constituents, quarks and gluons, 
two quarks tends to zero. This means that to be calculable within a 
perturbative approach, the interaction between these constituents
requires the presence in the process of a ``hard''
scale, i.e. a large transverse momentum transfer or a large mass.

The lepton beam of the high energy $e-p$ collider 
HERA is a prolific source of photon in a large virtuality range 
such that the study of $\gamma^* p$ interaction provides a completely new 
and deep insight into the QCD dynamics.

A major discovery at HERA is 
the observation of the strong rise of 
the total cross section at high energy in the deep inelastic scattering (DIS),
i.e. $\gamma^* p$ interactions with large $Q^2$ values
(\qsq\ being the negative of the squared four-momentum of the exchanged photon).
This is inconsistent with the case of the photoproduction ($Q^2 \simeq 0$),
which shows a soft dependence in the total hadronic 
energy, $W$, similar to the hadron-hadron interaction case~\cite{ref:caldwell}
and well described by the Regge phenomenological theory.
After a transition around $Q^2=$ 1 GeV$^2$,
the steep energy dependence of the total cross section in DIS 
is related to the fast increase
of the gluon density in the proton at high energy~\cite{ref:caldwell}.

HERA is thus a unique device to test QCD in the perturbative 
regime and to study the transition between perturbative and non-perturbative
domains. 
One of the remarkable success of this theory, as reviewed at this
conference~\cite{ref:caldwell}, 
is the correct prediction of the evolution of
the proton structure function with $Q^2$, for $Q^2 > 1$~GeV$^2$. 
This evolution allows the extraction of the gluon density in the 
proton, which is not directly measurable. 

The other main opened window at HERA for the understanding of the strong force
is the study of diffractive interaction.
Diffraction has been successfully described, 
already more than 30 years 
ago, via the introduction of an exchanged object carrying the vacuum quantum 
numbers, called the pomeron ($\Pom$).
Whilst Regge-based models give a unified description of all pre-HERA
diffractive data, this approach is not linked to the underlying
QCD theory.

\begin{figure}[htb] \unitlength 1mm
 \vspace{-1.0cm}
 \begin{center}
  \epsfig{file=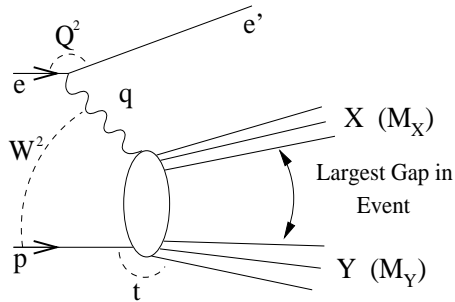,width=4cm,height=6cm,angle=270}
 \end{center}
 \vspace*{-1.2cm}
 \caption{Sketch of the diffractive $e-p$ interaction}
\label{fig:diag}
\end{figure}

The second major result at HERA is thus the observation in deep inelastic 
scattering that $8 - 10 \%$ of the events present a 
large rapidity gap (LRG) without hadronic activity between the 
two hadronic sub-systems, $X$ and $Y$, as illustrated in Fig.~\ref{fig:diag}.
\cite{diffr_1992}.
The gaps being significantly larger than implied by particle  
density fluctuation during the hadronisation process, these events
 are attributed 
to diffraction, i.e. to the exchange of a colourless object at the 
proton vertex.

\section{Exclusive vector meson production}
\begin{figure}[htb] \unitlength 1mm
 \vspace{1.3cm}
 \begin{picture}(100,100)(0,0) 
  \put(-2,-10){\epsfig{file=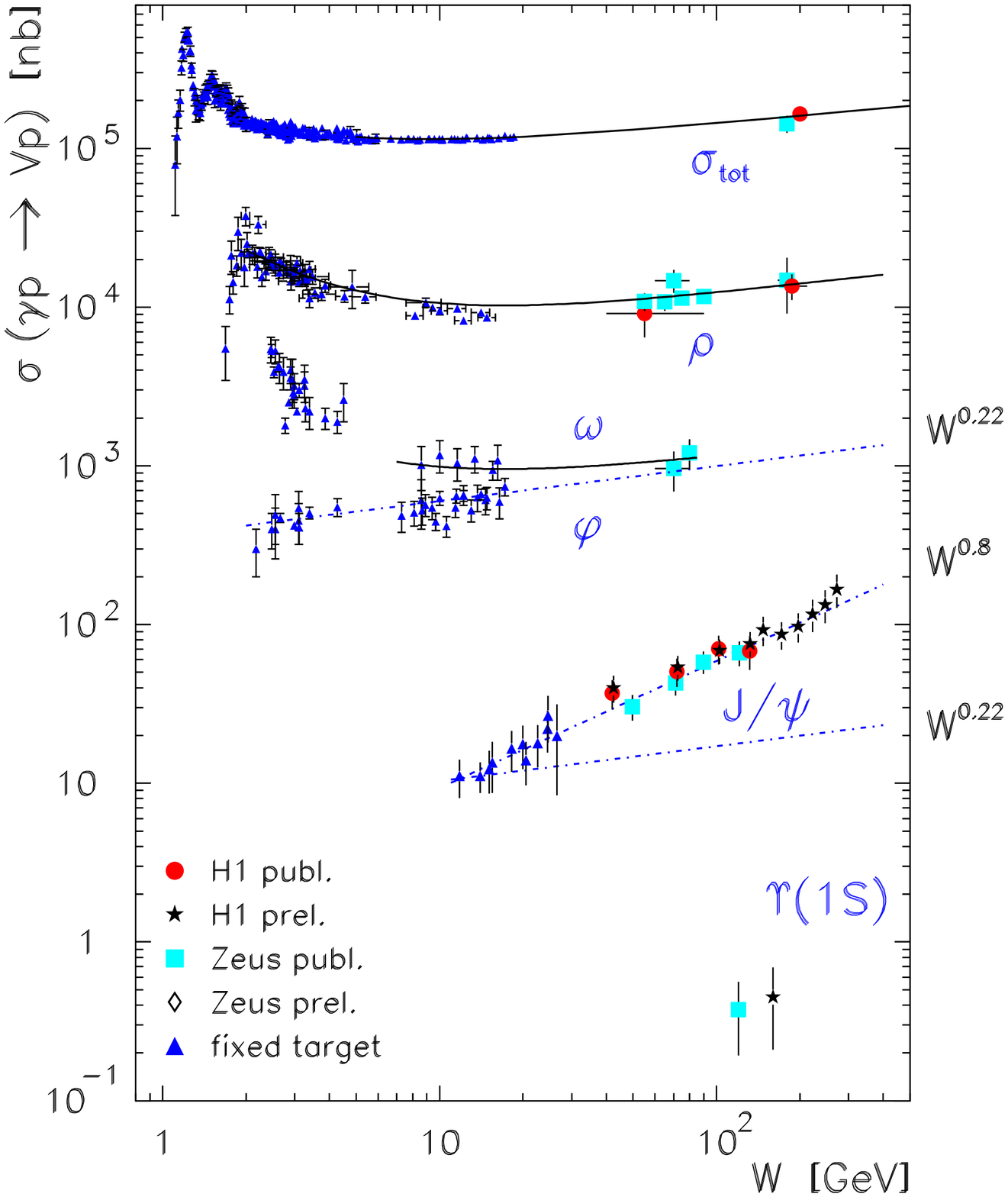,width=9.4cm,height=14cm}}
  \put(90,53){\epsfig{file=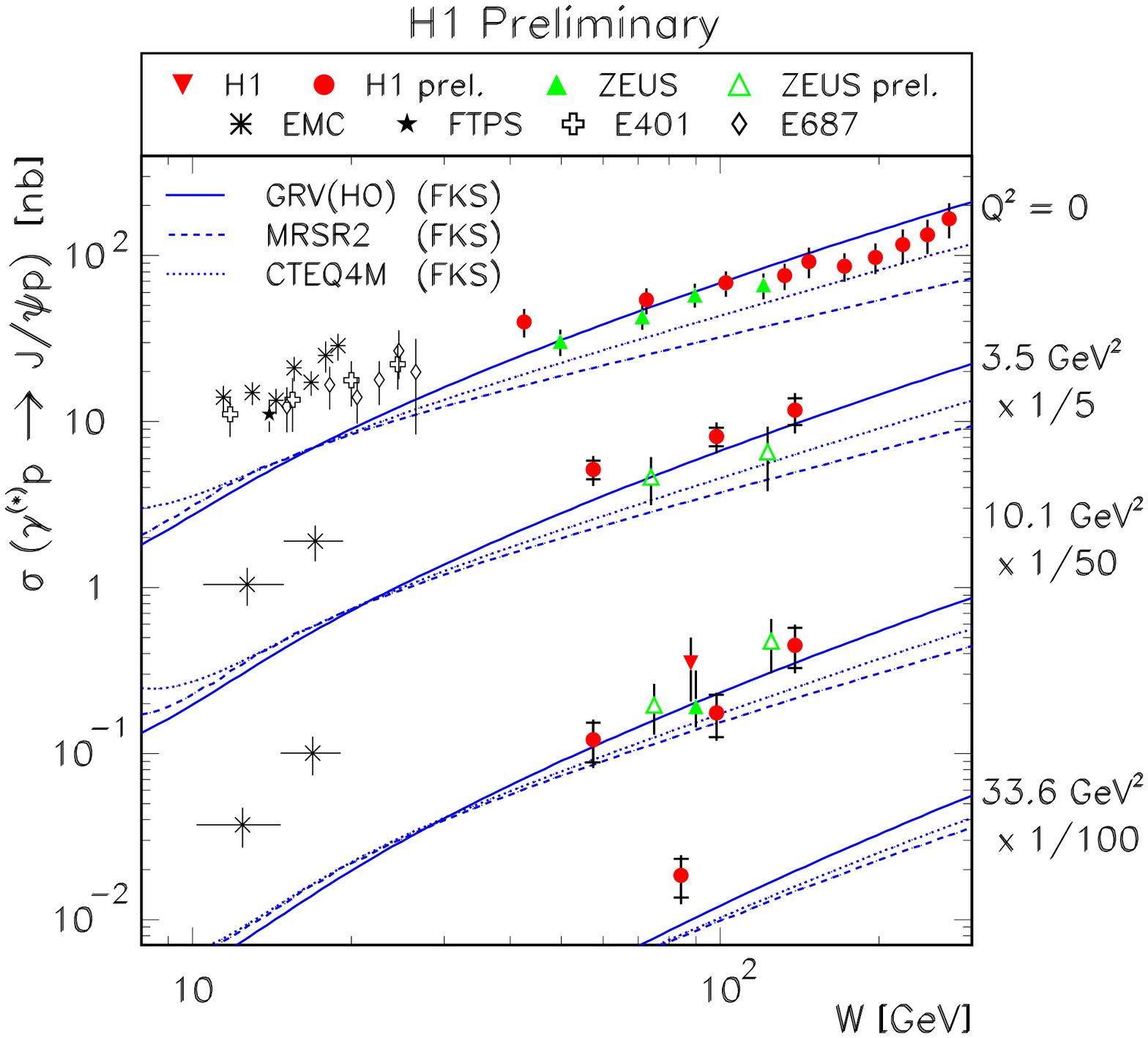,width=7.0cm,height=7cm}}
  \put(94,-6){\epsfig{file=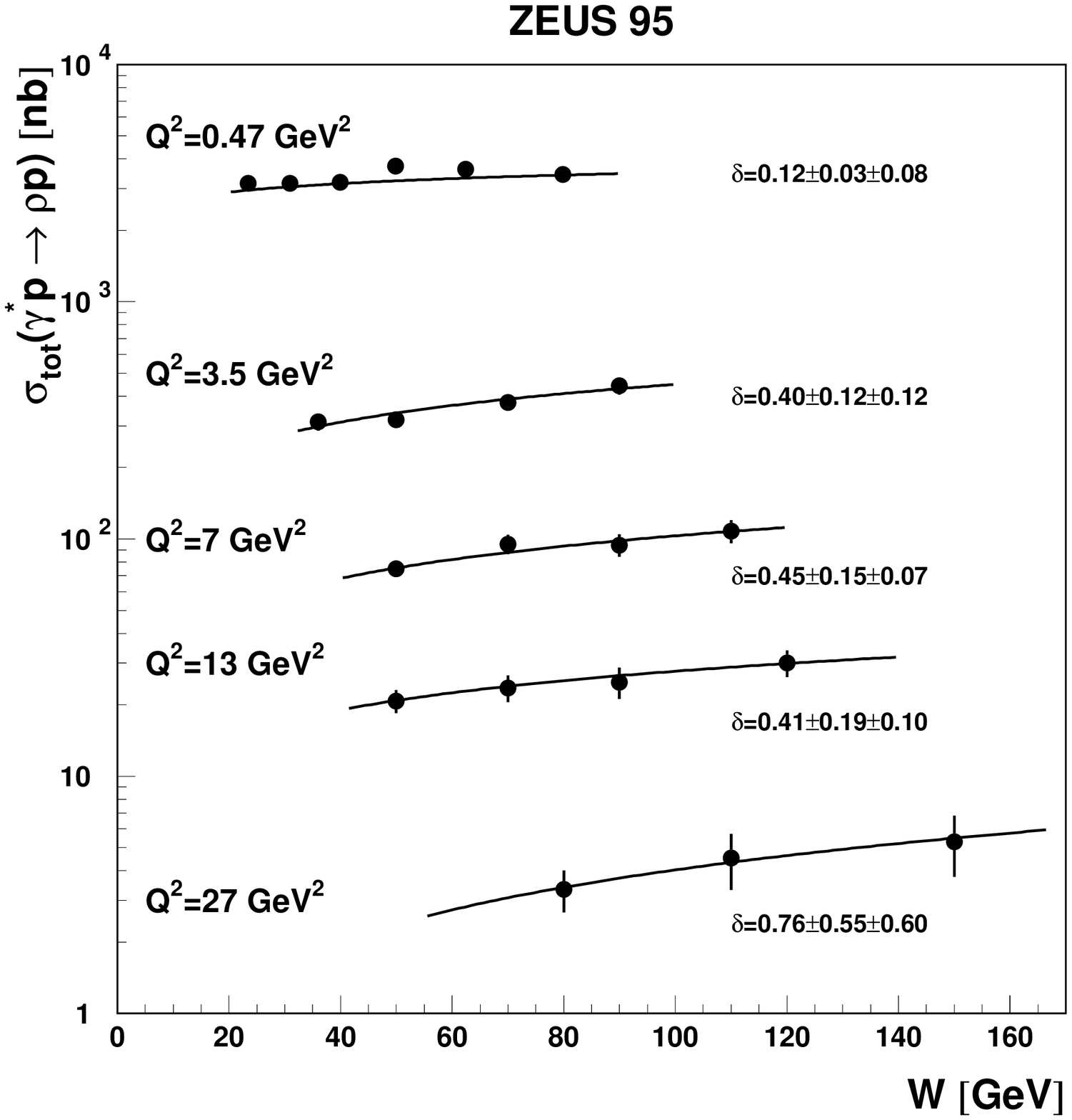,width=6.2cm,height=6.4cm}}
  \put(77,6){\bf a)}
  \put(144,64){\bf b)}
  \put(145,3){\bf c)}
 \end{picture}
 \vspace{-0.5cm}
 \caption{Diffractive cross sections as a function of the $\gamma^*-p$
system energy for: a) various vector mesons in photoproduction together 
with the total photoproduction cross section,
b) $J/\Psi$ in photoproduction and DIS, and 
c) $\rho$ in DIS for different $Q^2$ values.}
\label{fig:vm}
\end{figure}

 In exclusive elastic vector meson production study, 
$\gamma^* p \rightarrow V p$, the hadronic system $X$ consists 
only in a vector meson ($\rho, \omega,...$), 
and the system $Y$ in the scattered proton.
This process provides a very 
interesting way to test the mechanism of diffraction and our 
understanding of the pomeron structure.
Fig.~\ref{fig:vm} summarizes the $W$ dependence of various 
elastic exclusive vector meson production. Fig.~\ref{fig:vm}.a) presents the 
exclusive $\rho, \omega, \phi,J/\Psi$ and $\Upsilon$
production in photoproduction~\cite{ref:lightvm,ref:heavyvm}, 
together with the total photoproduction cross section~\cite{ref:totsig}. 
The light mesons ($\rho, \omega$ and $\phi$) show a soft dependence in $W$, 
equivalent to that of the total cross section dependence, 
while this energy dependence is much steeper
for $J/\Psi$ production.
This is interpreted as being due to the presence of a hard scale, the charm 
quark mass, making the $J/\Psi$ meson smaller than the
confinement scale ($\sim 1 fm$). 
In this case, it is natural to attempt a perturbative QCD description
of the process,
where the photon fluctuates into a quark-antiquark pair
and the exchanged pomeron is modeled by a pair of gluons.
This leads to a cross section proportional to the gluon 
density squared, which is in good agreement
(full line) with the data (points) shown on 
Fig.~\ref{fig:vm}.b)~\cite{ref:jpsi}.
This figure also shows the agreement of the 2 gluons exchange model
with measurement of exclusive  $J/\Psi$ production in the DIS regime, 
where a second hard scale, \qsq is present.

 As illustrated on Fig.~\ref{fig:vm}.c), a modification of the $W$ 
dependence also occurs for the elastic $\rho$ production
when the $Q^2$ increases. 

\section{Inclusive DIS cross section and partonic structure of the pomeron}

\begin{figure}[htb] \unitlength 1mm
 \vspace{-4.2cm}
 \begin{center}
 \begin{picture}(100,100)(0,0) 
   \put(-32,-10){\epsfig{file=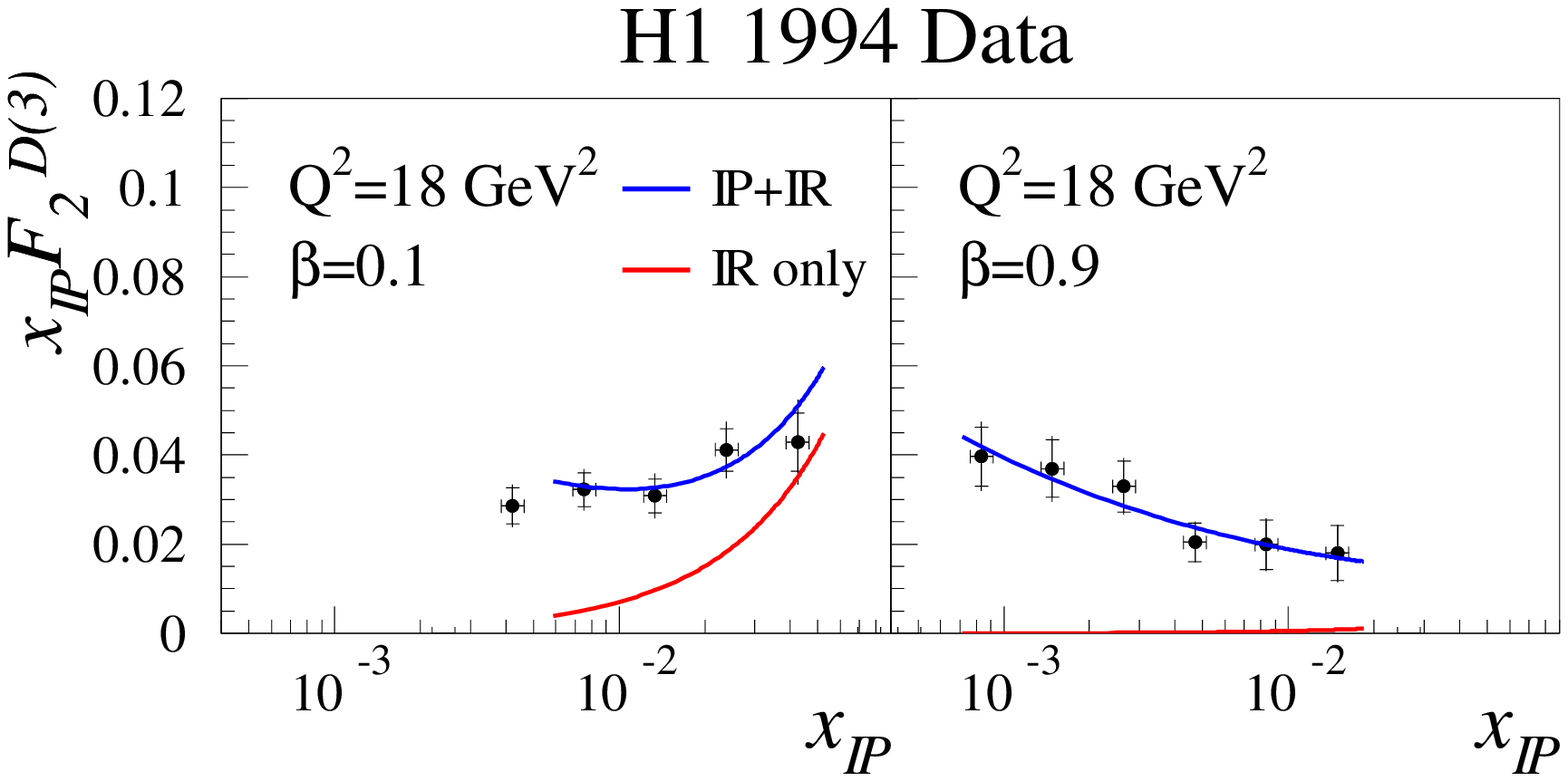,width=8.6cm,height=7cm}}
   \put(50,-13){\epsfig{file=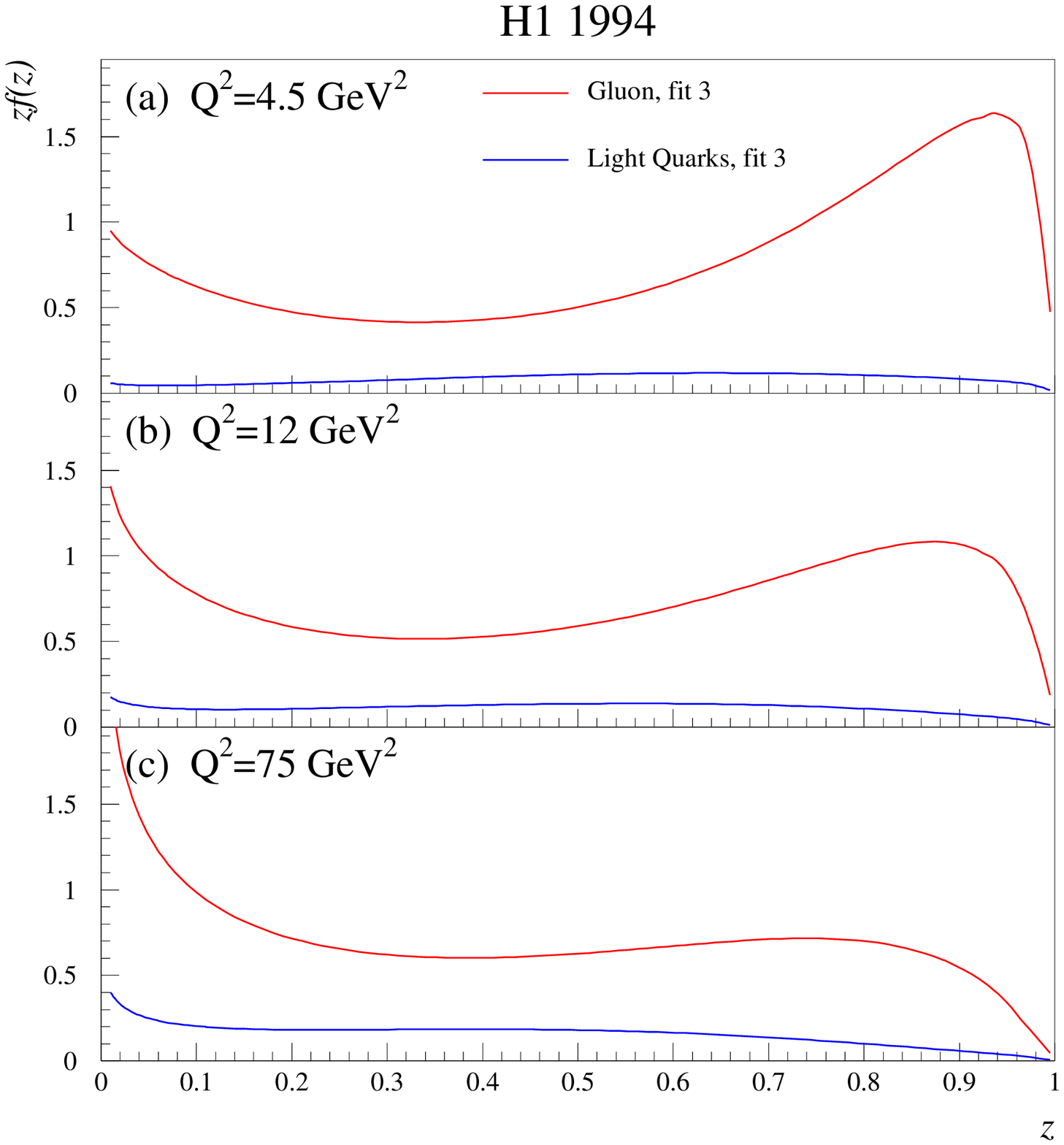,width=8cm,height=8.5cm}}
   \put(43,7){\bf a)}
   \put(123,12){\bf b)}
 \end{picture}
 \end{center}
 \vspace{-0.5cm}
 \caption{
a) Diffractive structure function measurement in two bins in $\beta$ and 
$Q^2$, as a function of $x_\Pom$.
Regge based phenomenological fit are overlaid with $\Pom$ and $\Reg$ or $\Reg$
only contributions.
b) Gluons and quark densities in the $\Pom$ extracted from the QCD fit of
\fdthree .}
\label{fig:f2d}
\end{figure}

The diffractive DIS process can be defined by four kinematic variables
conveniently chosen as  \qsq , \xpom , $\beta$ and \ttt, where 
\ttt \ is the squared
four-momentum transfer to the proton, and \xpom \ and $\beta$ are defined as
\begin{center}
\begin{equation}
   x_\Pom \simeq \frac{Q^2 + M_X^2}{Q^2 + W^2 }
   \qquad
    \beta \simeq \frac{Q^2 }{Q^2 + M_X^2} ; \label{eq:kin}
\end{equation}
\end{center}
\xpom\ can be interpreted as the fraction of the 
proton momentum carried by the exchanged
pomeron and $\beta $ is the fraction of the exchanged momentum carried by
the quark struck by the photon.
These variables are related to the Bjorken $x$ scaling variable (with
$W^2 \simeq Q^2 / x - Q^2$) by the relation
$ x = \beta \cdot x_\Pom$.

Experimentally, the $t$ variable is usually not measured or is integrated over.
In analogy with non-diffractive DIS scattering, the measured cross section
is expressed in the form of a three-fold diffractive structure function
\fdthreef :
\begin{equation}
  \frac { {\rm d}^3 \sigma \ (e p \rightarrow e X Y) }
 { {\rm d}Q^2 \ {\rm d}x_{I\!\!P} \ {\rm d}\beta}
        = \frac {4 \pi \alpha^2} {\beta Q^4}
            \ (1 - y + \frac {y^2}{2} )
            \ F_2^{D(3)} (Q^2, x_{I\!\!P} , \beta) ,
                                            \label{eq:fdthreefull}
\end{equation}
where $y$ is the usual scaling variable, with
$  y \simeq W^2 / s$.
\fdthree \ is conveniently factorised in the form
$F_2^{D(3)} (Q^2, x_{I\!\!P} , \beta ) =
     f_{\Pom/p} (x_{I\!\!P} ) \cdot F_2^D (Q^2, \beta)$,
assuming that the $\Pom$ flux $ f_{\Pom/p} (x_{I\!\!P})$ is independent
of the  $\Pom$ structure $F_2^D (Q^2, \beta)$, 
by analogy with the hadron structure functions, $\beta$ playing the role 
of Bjorken $x$. The  $\Pom$ flux is parametrized in a Regge inspired
form.
The fit of HERA data according to the Regge form has shown 
that factorization is broken.
This feature is explained in Regge theory by the need to include 
sub-leading trajectories in addition to the $\Pom$.
Including one further trajectory, the reggeon
($\Reg$), in addition to the pomeron: 
\begin{equation}
F_2^{D(3)} \ (Q^2, x_\Pom , \beta )
         =  f_{\Pom/p} (x_\Pom) \cdot F_2^\Pom (Q^2,\beta) +
              f_{\Reg/p} (x_\Pom) \cdot  F_2^\Reg (Q^2,\beta),
\end{equation}
is sufficient to obtain a good description of the data throughout
the measured kinematic domain ($0.4<Q^2<800$ GeV$^2$ $x_\Pom < 0.05$
and $0.001<\beta<0.9$)~\cite{ref:f2d}.

The contributions of pomeron and reggeon exchange are illustrated
on Fig.~\ref{fig:f2d}.a).
The reggeon contribution gets larger for increasing values of
\xpom, which correspond to smaller energy (for given \qsq\ and $\beta$
values).
It gets also larger for smaller values of $\beta$, which is consistent with
the expected decrease with $\beta$ of the reggeon structure function, following
the meson example, whereas the pomeron structure function is observed to
be approximately flat in $\beta$.
 
 By analogy to the QCD evolution of the proton structure function, 
one can attempt to extract the partonic structure of the pomeron
from the \qsq \ evolution of \fdtwo. The extracted distributions 
are shown in Fig.~\ref{fig:f2d}.b)
\cite{ref:f2d} separately for the gluon 
and the quark components as a function of $z$, the pomeron
momentum fraction carried by the parton entering the hard interaction.
This distribution shows the dominance of hard gluons (high $z$ values)
in the pomeron partonic structure.

 The dominance of hard gluons into the pomeron has been confirmed by various
analysis of the diffractive hadronic final 
state (jet production, energy flow, particle spectra and multiplicities,
and event shape) providing a global consistent picture of 
diffraction~\cite{ref:hadfin,ref:mult,ref:jets}.

\section*{Conclusion}

HERA experiments have produced a large amount of results in diffraction,
which allow confrontations with QCD predictions,
when one of the hard scales \qsq, the quark mass or \ttt \ 
(not reported in this summary) is present in the process.

 For the case of exclusive vector meson production, in the presence of a 
hard scale, models based on
the fluctuation of the photon in a quark-antiquark pair which
subsequently exchange a pair of gluons with the proton parton successfully 
reproduce the enhanced energy dependence.

 The QCD analysis of the total diffractive cross section, 
assuming factorization
into a pomeron flux in the proton  the corresponding parton distributions,
 favors the dominance of
hard gluons in the pomeron, confirmed by the analysis
 of inclusive final states and of jet production.

\end{document}